\documentstyle[12pt]{article}

\begin{document}
\title{New Dynamical Symmetry Breaking in Electroweak Theory}
\author{Bing An Li\\
Department of Physics and Astronomy, University of Kentucky\\
Lexington, KY 40506, USA}

\maketitle
\begin{abstract}
A new dynamical symmetry breaking of $SU(2)_{L}\times U(1)$
caused by the combination of
the axial-vector component and the fermion mass is found
in electroweak theory.
The masses of the W and
the Z bosons are obtained to be \(m^{2}_{W}={1\over2}g^{2}m^{2}_{t}\)
and \(m^{2}_{Z}=\rho m^{2}_{W}/cos^{2}\theta_{W}\) with
\(\rho\simeq 1\). They are in excellent agreement with data.
The Fermi constant is determined to be \(G_{F}={1\over 2\sqrt{2}m^{2}_{t}}\).
Two fixed gauge
fixing terms for W and Z boson fields are dynamically generated.
Massive neutrinos are required.
\end{abstract}

\newpage
The standard model[1] of electroweak interactions is
successful in many aspects. In this model Higgs is introduced to
generate the masses for the W, Z and
fermions by spontaneous symmetry breaking.
So far the experimental indication of the existence of the
Higgs has not been found yet. On the other hand, a
Higgs/Hierarchy problem[2] has been revealed.
There are many different attempts[3,4]
trying to solve this problem: W and Z bosons are composite; Higgs
fields are bound states of fermions; supersymmetry.
The top quark has been discovered in Fermi laboratory[5], whose mass
has been determined to be
\begin{equation}
m_{t}=180\pm 12 GeV [6].
\end{equation}
The value of $m_{t}$ is at the same order of magnitude
as the masses of the W and the
Z bosons. As a matter of fact, before the discovery of the top quark
there were attempts of finding the relationship between top quark and
intermediate bosons by using various mechanism[4].
In this paper a new approach is proposed to eliminate the Higgs and keep
the successes of the standard model.

This paper is organized as: 1) Lagrangian and formalism; 2) model of
new symmetry breaking; 3) masses of W and Z bosons; 4)propagators of W and Z
bosons;
5) Gauge fixing term;
6) theoretical values of $m_{W}$ and $m_{Z}$; 7)summary.
\section{Lagrangian and formalism}
The Lagrangian of the standard model consists of boson fields,
fermions, and Higgs.
The couplings between fermions and bosons have been extensively tested.
Theoretical results are in excellent agreement with data.
The mass
terms of fermions(except for neutrinos) are well established.
The Lagrangian of the boson fields is constructed by gauge principle.
Therefore,
in the Lagrangian of the standard model
the part of the boson fields, the interactions
between fermions and bosons, and the mass terms of fermions are
reliable.
On the other hand,
the Higgs sector of the Lagrangian of the standard model
has not been determined yet. In this paper we study the dynamical
properties of the Lagrangian without Higgs
\begin{eqnarray}
\lefteqn{{\cal L}=
-{1\over4}A^{i}_{\mu\nu}A^{i\mu\nu}-{1\over4}B_{\mu\nu}B^{\mu\nu}
+\bar{q}\{i\gamma\cdot\partial-M\}q}
\nonumber \\
&&+\bar{q}_{L}\{{g\over2}\tau_{i}
\gamma\cdot A^{i}+g'{Y\over2}\gamma\cdot B\}
q_{L}+\bar{q}_{R}g'{Y\over2}\gamma\cdot Bq_{R}\nonumber \\
&&+\bar{l}\{i\gamma\cdot\partial-M_{f}\}l
+\bar{l}_{L}\{{g\over2}
\tau_{i}\gamma\cdot A^{i}-{g'\over2}\gamma\cdot B\}
l_{L}-\bar{l}_{R}g'\gamma\cdot B l_{R}.
\end{eqnarray}
Summation over $q_{L}$, $q_{R}$, $l_{L}$, and $L_{R}$ is
implicated in Eq.(2).

It is necessary to point out that in eq.(2) the boson fields are still
elementary fields and
the couplings
between the bosons and the fermions of the standard model
remain unchanged, therefore, the successes of the standard model are
kept.
In the standard model the masses of the W and Z bosons are via
spontaneous symmetry breaking mechanism generated by Higgs. In this
paper we study whether $m_{W}$ and $m_{Z}$ can be dynamically
generated from this Lagrangian(2).

Due to
the fermion mass terms the Lagrangian(2) is no longer gauge invariant.
Without losing generality, we study the properties of the
Lagrangaian of the
generation of t and b quarks. The Lagrangian
of this doublet is
\begin{eqnarray}
\lefteqn{{\cal L}=
-{1\over4}A^{i}_{\mu\nu}A^{i\mu\nu}-{1\over4}B_{\mu\nu}B^{\mu\nu}
+\bar{t}\{i\gamma\cdot\partial-m_{t}\}t
+\bar{b}\{i\gamma\cdot\partial-m_{b}\}b}\nonumber \\
&&+\bar{\psi}_{L}\{{g\over2}\tau_{i}
\gamma\cdot A^{i}+g'{1\over6}\gamma\cdot B\}
\psi_{L}
+{2\over3}g'\bar{t}_{R}\gamma\cdot Bt_{R}
-{1\over3}g'\bar{b}_{R}\gamma
\cdot Bb_{R},
\end{eqnarray}
where
\(\psi_{L}=\left(\begin{array}{c}
                t\\b
                 \end{array} \right)_{L}.\)
The quark part of
the Lagrangian(3) is rewritten as
\begin{equation}
{\cal L}=\bar{\psi}\{i\gamma\cdot\partial+\gamma\cdot v+
\gamma\cdot a\gamma_{5}-m\}\psi,
\end{equation}
where $\psi$ is the doublet of t and b quarks,
\(m=\left(\begin{array}{c}
          m_{t}\hspace{1cm}0\\
          0\hspace{1cm}m_{b}
          \end{array} \right),\)
\(v_{\mu}=\tau_{i}v^{i}_{\mu}+\omega_{\mu}\),
\(v^{1,2}_{\mu}={g\over4}A^{1,2}_{\mu}\),
\(v^{3}_{\mu}={g\over4}A^{3}_{\mu}+{g'\over4}B_{\mu}\),
\(\omega_{\mu}={g'\over6}B_{\mu}\),
\(a_{\mu}=\tau_{i}a^{i}_{\mu}\),
\(a^{1,2}_{\mu}=-{g\over4}
A^{1,2}_{\mu}\),
\(a^{3}_{\mu}=-{g\over4}A^{3}_{\mu}+{g'\over4}B_{\mu}\).

The mass term, $\bar{\psi}m\psi$, is not invariant under the
transformation
\[\psi\rightarrow e^{i(\alpha_{1}\tau_{1}+\alpha_{2}
\tau_{2})}\psi.\]
Therefore, the charged gauge symmetry is explicitly broken by the quark
masses and
the charged bosons, W, are expected to gain masses.
On the other hand, the quark mass term is invariant under two gauge
transformations
\[\psi\rightarrow e^{i\alpha}\psi\]
and
\[\psi\rightarrow e^{i\alpha_{3}\tau_{3}}\psi.\]
The theory should have two massless neutral bosons. One of the two
neutral bosons is photon. It is obvious that a new dynamical symmetry
is needed for making the Z boson massive.

Because the Lagrangian(4) is not invariant
under the charged gauge transformation in principle the
mass term of the W boson can be added to the Lagrangian(2). This
paper shows that the mass of the W boson can be correctly generated
by the Lagrangian(2). Therefore, the mass term of W boson is fine-tuned
to be zero. This kind of phenomena is not only for gauge symmetry broken
theory, it happens to gauge invariant theory too. For example, in QED
there are many other gauge invariant terms which are not forbidden
by any physical principle.
Anomalous magnetic moment of the lepton is a good example and it
can be added to the
Lagrangian. However, it is well known that the anomalous magnetic moment
of the lepton is derived from the loop diagrams. There is no need to
put such term into the Lagrangian by hand. This is known as the minimum
coupling principle.

Using path integral
to integrate out the quark fields, in Euclidean space
the Lagrangian of boson fields
is obtained
\begin{equation}
{\cal L}=lndetD,
\end{equation}
where
\[{\cal D}=\gamma\cdot\partial-i\gamma\cdot v-i\gamma\cdot a\gamma_{5}+m.\]
The real and imaginary parts of the Lagrangian(5) are
\begin{equation}
{\cal L}_{Re}={1\over2}ln det({\cal D}^{\dag}{\cal D}),\;\;\;
{\cal L}_{Im}={1\over2}ln det(\frac{{\cal D}}{{\cal D}^{\dag}}),
\end{equation}
where
\begin{equation}
{\cal D}^{\dag}=-
\gamma\cdot\partial+i\gamma\cdot v-i\gamma\cdot a\gamma_{5}+m.
\end{equation}

It is necessary to point out that
${\cal D}^{\dag}{\cal D}$ is a definite positive operator.
In terms of Schwinger's proper time method[7] ${\cal L}_{Re}$ is
expressed as
\begin{equation}
{\cal L}_{Re}={1\over2}\int d^{D}xTr\int^{\infty}_{0}{d\tau\over\tau}
e^{-\tau{\cal D}^{\dag}{\cal D}}.
\end{equation}
Inserting a complete set of plane waves and subtracting the divergence
at \(\tau=0\), we obtain
\begin{equation}
{\cal L}_{Re}={1\over2}\int d^{D}x
\frac{d^{D}p}{(2\pi)^{D}}Tr\int^{\infty}_{0}{d\tau\over \tau}
\{e^{-\tau{\cal D'}^{\dag}{\cal D}^{'}}-e^{-\tau\Delta_{0}}\},
\end{equation}
where
\begin{eqnarray}
\lefteqn{{\cal D'}=\gamma\cdot\partial+i\gamma\cdot p
-i\gamma\cdot v-i\gamma
\cdot a\gamma_{5}+m,\;\;\;
{\cal D}'^{\dag}=-\gamma\cdot\partial-i\gamma\cdot p+i\gamma\cdot v
-i\gamma\cdot a\gamma_{5}+m,}\nonumber \\
&&{\cal D}'^{\dag}{\cal D}'=\Delta_{0}-\Delta,\;\;\;
\Delta_{0}=p^{2}+m^{2}_{1},\;\;\;
m^{2}_{1}={1\over2}(m^{2}_{t}+m^{2}_{b}),\;\;\;
m^{2}_{2}={1\over2}(m^{2}_{t}-m^{2}_{b}),\nonumber \\
&&\Delta=\partial^{2}-(\gamma\cdot v-\gamma\cdot a\gamma_{5})
(\gamma\cdot v+\gamma\cdot a\gamma_{5})-i\gamma\cdot\partial
(\gamma\cdot v+\gamma\cdot a\gamma_{5})\nonumber \\
&&-i(\gamma\cdot v-\gamma\cdot a
\gamma_{5})\gamma\cdot\partial+2ip\cdot\partial
+2p\cdot(v+a\gamma_{5})
-i[\gamma\cdot v,m]\nonumber \\
&&+i\{\gamma\cdot a,m\}\gamma_{5}
-m^{2}_{2}\tau_{3}.
\end{eqnarray}
After the integration over $\tau$, ${\cal L}_{Re}$ is expressed as
\begin{equation}
{\cal L}_{Re}={1\over2}\int d^{D}x\frac{d^{D}p}{(2\pi)^{D}}
\sum^{\infty}_{n=1}{1\over n}\frac{1}{(p^{2}+m^{2}_{1})^{n}}
Tr\Delta^{n}.
\end{equation}
Due to \({\cal L}_{IM}(-m,-\gamma_{5})=-{\cal L}_{IM}(m,\gamma_{5})\),
${\cal L}_{Im}$ doesn't contribute to the masses of W and Z bosons
at least at the tree level of the boson fields. Therefore, the study
of ${\cal L}_{Im}$ is beyond the scope of this paper..

${\cal L}_{Re}$ is used to investigate
the symmetry breaking mechanism, the masses of the intermediate bosons,
and the propagators of the boson fields.
As mentioned above,
the $SU(2)_{L}\times U(1)$ symmetry is explicitly broken by both
quark and lepton masses.
However, another symmetry breaking mechanism is needed.

It is necessary to emphasize on that the field theory of electroweak
interactions is different from $QED$ and $QCD$. In $QED$ and $QCD$
photon and gluons are pure vector fields.
{\bf Due to parity
nonconservation the intermediate bosons in the standard model
have both  vector and
axial-vector components which are written in the forms of
v and a in
Eq.(4).}
In this paper it is shown that this property of the intermediate
boson fields results in another
$SU(2)_{L}\times U(1)$ symmetry breaking.
From the expression of $\Delta$(10)
it is seen that in company with fermion mass
the vector component v of the intermediate boson field
appears in a commutator $[v,m]$, while the axial-vector
component a appears in an anticommutator
$\{a,m\}$. Due to this property
the axial-vector component of boson field causes a new
symmetry breaking.
\section{Model of new symmetry breaking}
In order to show how the axial-vector field results in a symmetry breaking
a model is studied in this section.
The Lagrangian of a vector field and a
fermion($QED$) is
\begin{equation}
{\cal L}=-{1\over4}F_{\mu\nu}F^{\mu\nu}+\bar{\psi}\{i\gamma\cdot
\partial+e\gamma\cdot v\}\psi-m\bar{\psi}\psi.
\end{equation}
This Lagrangian(12) is invariant under the gauge transformation
\[\psi\rightarrow e^{i\alpha(x)}\psi,\;\;\; v_{\mu}\rightarrow
v_{\mu}+{1\over e}\partial_{\mu}\alpha.\]
Using the Eqs.(10,11), the mass term of the vector field is obtained
\begin{equation}
{\cal L}_{M}={1\over2}\int d^{D}x\int\frac{d^{D}p}{(2\pi)^{D}}
\sum^{\infty}_{n=1}{1\over n}\frac{1}{(p^{2}+m^{2})^{n}}
Tr(2p\cdot v-v^{2})^{n}.
\end{equation}
Only \(n=1, 2\) contribute to the mass term
\begin{equation}
{\cal L}_{M}=-{D\over2}\int d^{D}x\int\frac{d^{D}p}{(2\pi)^{D}}
\frac{1}{p^{2}+m^{2}}v^{2}+D\int d^{D}x\int\frac{d^{D}p}{(2\pi)^{D}}
\frac{1}
{(p^{2}+m^{2})^{2}}p\cdot vp\cdot v=0.
\end{equation}
These two terms cancel each other. As expected,
no mass is generated.

The Lagrangian of an axial-vector field and a
fermion is
\begin{equation}
{\cal L}=-{1\over4}F_{\mu\nu}F^{\mu\nu}+\bar{\psi}\{i\gamma\cdot
\partial+e\gamma\cdot a\gamma_{5}\}\psi-m\bar{\psi}\psi.
\end{equation}
This Lagrangian(15) is invariant under the gauge transformation
\[\psi\rightarrow e^{i\alpha(x)}\psi,\;\;\; a_{\mu}\rightarrow
a_{\mu}+{1\over e}\partial_{\mu}\alpha\gamma_{5}.\]
Using the Eqs.(10,11), the mass term of the
axial-vector field is obtained
\begin{equation}
{\cal L}_{M}={1\over2}\int d^{D}x\int\frac{d^{D}p}{(2\pi)^{D}}
\sum^{\infty}_{n=1}{1\over n}\frac{1}{(p^{2}+m^{2})^{n}}
Tr(2p\cdot a\gamma_{5}-a^{2}+i\{\gamma\cdot a,m\}\gamma_{5})^{n}.
\end{equation}
Only \(n=1, 2\) contribute to the mass term
\begin{eqnarray}
{\cal L}_{M}=-{D\over2}\int d^{D}x\int\frac{d^{D}p}{(2\pi)^{D}}
\frac{1}{p^{2}+m^{2}}a^{2}+D\int d^{D}x\int\frac{d^{D}p}{(2\pi)^{D}}
\frac{1}
{(p^{2}+m^{2})^{2}}\{p\cdot ap\cdot a+m^{2}a^{2}\}\nonumber \\
=\frac{1}{(4\pi)^{2}}D\Gamma(2-{D\over2})m^{2}a_{\mu}a^{\mu}.
\end{eqnarray}
The axial-vector field gains mass, therefore, the symmetry is broken.

Comparing with eq.(13), in Eq.(16) there is one more term
which is the anticommutator
$i\{\gamma\cdot a, m\}\gamma_{5}$. Because of \([v_{\mu}, m]=0\)
there is no such term for vector field. Therefore, the theory of
axial-vector field is very different from the vector field.
The symmetry in the theory
of axial-vector field is broken by the combination of the axial-vector
field and the mass of the fermion. The mass of the axial-vector field
is generated by this symmetry breaking.
\section{Masses of W and Z bosons}
In terms of the Lagrangian(11) the masses of the
intermediate bosons are calculated.
The terms related to the masses only
is separated from Eq.(11)
\begin{eqnarray}
\lefteqn{{\cal L}_{M}={1\over2}\int d^{D}x\int\frac{d^{D}p}{(2\pi)^{D}}
\sum^{\infty}_{n=1}{1\over n}\frac{1}{(p^{2}+m^{2}_{1})^{n}}Tr\{
-(\gamma\cdot v-\gamma\cdot a\gamma_{5})}\nonumber \\
&&(\gamma\cdot v+\gamma\cdot
a\gamma_{5})+2p\cdot(v+a\gamma_{5})
+i[m,\gamma\cdot v]+i\{m,
\gamma\cdot a\}\gamma_{5}-m^{2}_{2}\tau_{3}\}^{n}.
\end{eqnarray}
The contributions of the fermion masses to $m_{W}$ and $m_{Z}$
are needed to be calculated to all orders.

Four kinds of terms of the Lagrangian(18)
contribute to the masses of the bosons
\begin{eqnarray}
\lefteqn{{\cal L}^{1}=-{1\over2}\int\frac{d^{D}p}{(2\pi)^{D}}\sum^{\infty}
_{n=1}{1\over n}
\frac{1}{(p^{2}+m^{2}_{1})^{n}}Tr(\gamma\cdot v-\gamma\cdot a\gamma_{5})
(\gamma\cdot v+\gamma\cdot a\gamma_{5})(-m^{2}_{2}\tau_{3})^{n-1},}\nonumber \\
&&{\cal L}^{2}=
2\int\frac{d^{D}p}{(2\pi)^{D}}\sum^{\infty}
_{n=2}{1\over n}\frac{(-m^{2}_{2})^{n-2}}{(p^{2}+m^{2}_{1})^{n}}
\sum^{n-2}_{k=0}(n-1-k)Tr
p\cdot(v+a\gamma_{5})\tau_{3}^{k}
p\cdot(v+a\gamma_{5})\tau_{3}^{n-2-k},\nonumber \\
&&{\cal L}^{1+2}={\cal L}^{1}+{\cal L}^{2}=-\frac{8N_{c}}{(4\pi)^{2}}
\sum^{\infty}_{k=1}\frac{1}{(2k+1)(2k-1)}(\frac{m^{2}_{2}}{m^{2}_{1}})
^{2k}({g\over4})^{2}m^{2}_{1}\sum^{2}_{i=1}A^{i}_{\mu}A^{i\mu},\\
&&{\cal L}^{3}=-{1\over2}\int\frac{d^{D}p}{(2\pi)^{D}}\sum^{\infty}
_{n=2}{1\over n}\frac{(-m^{2}_{2})^{n-2}}{(p^{2}+m^{2}_{1})^{n}}\sum^{n-2}
_{k=0}(n-1-k)Tr[\gamma\cdot v,m]\tau_{3}^{k}[\gamma\cdot
v,m]\tau_{3}^{n-2-k}\nonumber \\
&&=\frac{2N_{c}}{(4\pi)^{2}}D\Gamma(2-{D\over2})({g\over4})^{2}
m^{2}_{-}\sum^{2}_{i=1}A^{i}_{\mu}A^{i\mu}\nonumber \\
&&+\frac{8N_{C}}{(4\pi)^{2}}\sum^{\infty}_{k=2}\frac{1}
{(2k-1)(2k-2)}(\frac{m^{2}_{2}}{m^{2}_{1}})^{2k-2}({g\over4})^{2}
m^{2}_{-}\sum^{2}_{i=1}A^{i}_{\mu}A^{i\mu},\\
&&{\cal L}^{4}={1\over2}\int\frac{d^{D}p}{(2\pi)^{D}}\sum^{\infty}
_{n=2}{1\over n}\frac{(m^{2}_{2})^{n-2}}{(p^{2}+m^{2}_{1})^{n}}\sum^{n-2}
_{k=0}(n-1-k)Tr\{\gamma\cdot a,m\}\tau_{3}^{k}
\{\gamma\cdot
a,m\}\tau_{3}^{n-2-k}\nonumber \\
&&=\frac{2N_{c}}{(4\pi)^{2}}D\Gamma(2-{D\over2})\{
({g\over4})^{2}
m^{2}_{+}\sum^{2}_{i=1}A^{i}_{\mu}A^{i\mu}
+m^{2}_{1}[({g\over4})^{2}+({g'\over4})^{2}]Z_{\mu}Z^{\mu}\}\nonumber \\
&&-\frac{8N_{C}}{(4\pi)^{2}}\sum^{\infty}_{k=1}\frac{1}
{(2k-1)2k}(\frac{m^{2}_{2}}{m^{2}_{1}})^{2k-1}\{({g\over4})^{2}
+({g'\over4})^{2}\}
m^{2}_{2}Z_{\mu}Z^{\mu}\nonumber \\
&&+\frac{8N_{C}}{(4\pi)^{2}}\sum^{\infty}_{k=1}\frac{1}
{(2k+1)2k}(\frac{m^{2}_{2}}{m^{2}_{1}})^{2k}({g\over4})^{2}
m^{2}_{+}\sum^{2}_{i=1}A^{i}_{\mu}A^{i\mu},
\end{eqnarray}
where \(m_{+}={1\over2}(m_{t}+m_{b})\) and \(m_{-}={1\over2}
(m_{t}-m_{b})\).

The Eqs.(19-21) show that all the four terms contribute to
the mass of the W boson.
These results indicate that the mass of the charged boson, W,
originates in both the explicit $SU(2)_{L}\times U(1)$
symmetry breaking by the fermion masses and the dynamical symmetry
breaking caused by the combination of the axial-vector component
and the fermion mass($\{a_{\mu},m\}$).
Only ${\cal L}^{4}$ contributes to $m_{Z}$. ${\cal L}^{4}$ is related to $
\{a_{\mu},m\}$. Therefore, $m_{Z}$ is dynamically generated by the new
dynamical symmetry breaking.
As expected, a U(1) symmetry
remains and the neutral vector meson, the photon,
is massless.

It is found that the series of the fermion masses are convergent to
analytic functions.
Putting all the four terms(19-21) together, the masses of W and Z
are obtained in Minkowski space
\begin{eqnarray}
\lefteqn{{\cal L}_{M}={1\over2}\frac{N_{C}}{(4\pi)^{2}}\{{D\over4}
\Gamma(2-{D\over2})(4\pi{\mu^{2}\over m^{2}_{1}})
^{{\epsilon\over2}}+{1\over2}[1-ln(1-x)-
(1+{1\over x}){\sqrt{x}\over2}
ln\frac{1+\sqrt{x}}{1-\sqrt{x}}]\}m^{2}_{1}g^{2}
\sum^{2}_{i=1}A^{i}_{\mu}A^{i\mu}}\nonumber \\
&&+{1\over2}\frac{N_{C}}{(4\pi)^{2}}\{{D\over4}
\Gamma(2-{D\over2})(4\pi{\mu^{2}\over m^{2}_{1}})
^{{\epsilon\over2}}-{1\over2}[ln(1-x)+\sqrt{x}
ln\frac{1+\sqrt{x}}{1-\sqrt{x}}]\}m^{2}_{1}(g^{2}+g'^{2})
Z_{\mu}Z^{\mu},
\end{eqnarray}
where $N_{C}$ is the number of colors and
\(x=({m^{2}_{2}\over m^{2}_{1}})^{2}\). It is necessary to point
out that
\begin{equation}
a^{3}_{\mu}={1\over4}\sqrt{g^{2}+g'^{2}}Z_{\mu}.
\end{equation}
In the same way, other two generations of quarks,
\(\left(\begin{array}{c}
         u\\d
        \end{array} \right) \)
and
\(\left(\begin{array}{c}
         c\\s
        \end{array} \right) \),
and three generations of leptons
\(\left(\begin{array}{c}
         \nu_{e}\\e
        \end{array} \right) \),
\(\left(\begin{array}{c}
         \nu_{\mu}\\\mu
        \end{array} \right) \), and
\(\left(\begin{array}{c}
         \nu_{\tau}\\\tau
        \end{array} \right) \)
contribute to the masses of W and Z bosons too. By changing the
definitions of $m^{2}_{1}$ and x to the quantities of other
generations in Eq.(22),
the contributions of the other two quark generations
are found. Taking off the factor $N_{C}$ and changing $m^{2}_{1}$ and
x to corresponding quantities of leptons, the contributions of the
leptons to $m_{W}$ and $m_{Z}$ are obtained.
In this paper the effects of CKM matrix are not taken into account.
The final expressions of the masses of W and Z bosons
are the sum of the contributions of the three quark and the three
lepton generations.
It is learned from the processes deriving Eq.(22) that
\begin{enumerate}
\item Due to the U(1) symmetry the neutral vector
field $sin\theta_{W}A^{3}_{\mu}+cos\theta_{W}B_{\mu}$(photon
field) is massless;
\item The W boson gains mass from both the explicit and dynamical
$SU(2)_{L}\times U(1)$ symmetry breaking;
\item
$m_{Z}$ is resulted in the dynamical symmetry only.
Without $\{a_{\mu}, m\}$ Z boson is massless.
\end{enumerate}
\section{Propagators of W and Z bosons}
In the standard model the bosons are Yang-Mills fields and massless
before the spontaneous symmetry breaking. Gauge fixing terms can
be chosen and the theory is renormalizable. In the Lagrangian(2)
the boson fields are still Yang-Mills fields. However after the
symmetry breaking is taken into account, as done above, the W and
the Z gain masses.
They are massive. It is necessary to study their
propagators to see whether they have right behavior for
renormalization at high energy.
Up to all orders of fermion masses,
the kinetic terms of the intermediate boson fields
are obtained from the Lagrangian(11). For the generation of t and b quarks
there are nine terms
\begin{eqnarray}
\lefteqn{{\cal L}^{1}=-{1\over2}\int\frac{d^{D}p}{(2\pi)^{D}}\sum^{\infty}
_{n=3}{1\over n}\frac{(-m^{2}_{2})^{n-3}}{(p^{2}+m^{2}_{1})^{n}}\sum
^{n-3}_{k=0}\sum^{n-3-k}_{k_{1}=0}(n-2-k-k_{1})Tr[\gamma\cdot v,m]
\partial^{2}\tau^{k+k_{1}}_{3}[\gamma\cdot v,m]\tau^{n-3-k-k_{1}}
}\nonumber \\
&&=\frac{8N_{C}}{(4\pi)^{2}}\frac{m^{2}_{-}}{m^{2}_{1}}\sum^{\infty}
_{k=1}\frac{1}{(2k+1)(2k-1)}(\frac{m^{2}_{2}}{m^{2}_{1}})^{2k-2}
({g\over4})^{2}\sum^{2}_{i=1}A^{i}_{\mu}\partial^{2}A^{i\mu},\\
&&{\cal L}^{2}=2\int\frac{d^{D}p}{(2\pi)^{D}}\sum^{\infty}
_{n=4}{1\over n}\frac{(-m^{2}_{2})^{n-4}}{(p^{2}+m^{2}_{1})^{n}}\sum
^{n-4}_{k=0}\sum^{n-4-k}_{k_{1}=0}\sum^{n-4-k-k_{1}}_{k_{2}=0}
(n-3-k-k_{1}-k_{2})\nonumber \\
&&Tr[\gamma\cdot v,m]\tau^{k+k_{1}+k_{2}}_{3}(p\cdot \partial)^{2}
[\gamma\cdot v,m]\tau^{n-4-k-k_{1}-k_{2}}
\nonumber \\
&&=-\frac{4N_{C}}{(4\pi)^{2}}\frac{m^{2}_{-}}{m^{2}_{1}}\sum^{\infty}
_{k=1}\frac{1}{(2k+1)(2k-1)}(\frac{m^{2}_{2}}{m^{2}_{1}})^{2k-2}
({g\over4})^{2}\sum^{2}_{i=1}A^{i}_{\mu}\partial^{2}A^{i\mu},\\
&&{\cal L}^{3}={1\over2}\int\frac{d^{D}p}{(2\pi)^{D}}\sum^{\infty}
_{n=3}{1\over n}\frac{(-m^{2}_{2})^{n-3}}{(p^{2}+m^{2}_{1})^{n}}\sum
^{n-3}_{k=0}\sum^{n-3-k}_{k_{1}=0}(n-2-k-k_{1})\nonumber \\
&&Tr\{\gamma\cdot a,m\}
\partial^{2}\tau^{k+k_{1}}_{3}\{\gamma\cdot a,m\}\tau^{n-3-k-k_{1}}
\nonumber \\
&&=\frac{8N_{C}}{(4\pi)^{2}}\frac{m^{2}_{+}}{m^{2}_{2}}\sum^{\infty}
_{k=1}\frac{1}{(2k+1)(2k-1)}(\frac{m^{2}_{2}}{m^{2}_{1}})^{2k-1}
({g\over4})^{2}\sum^{2}_{i=1}A^{i}_{\mu}\partial^{2}A^{i\mu}+
\frac{8N_{C}}{(4\pi)^{2}}{1\over3}a^{3}_{\mu}\partial^{2}a^{3\mu},\\
&&{\cal L}^{4}=-2\int\frac{d^{D}p}{(2\pi)^{D}}\sum^{\infty}
_{n=4}{1\over n}\frac{(-m^{2}_{2})^{n-4}}{(p^{2}+m^{2}_{1})^{n}}\sum
^{n-4}_{k=0}\sum^{n-4-k}_{k_{1}=0}\sum^{n-4-k-k_{1}}_{k_{2}=0}
(n-3-k-k_{1}-k_{2})\nonumber \\
&&Tr\{\gamma\cdot a,m\}\tau^{k+k_{1}+k_{2}}_{3}(p\cdot \partial)^{2}
\{\gamma\cdot a,m\}\tau^{n-4-k-k_{1}-k_{2}}
\nonumber \\
&&=-\frac{4N_{C}}{(4\pi)^{2}}\frac{m^{2}_{+}}{m^{2}_{2}}\sum^{\infty}
_{k=2}\frac{1}{(2k-1)(2k-3)}(\frac{m^{2}_{2}}{m^{2}_{1}})^{2k-3}
({g\over4})^{2}\sum^{2}_{i=1}A^{i}_{\mu}\partial^{2}A^{i\mu}
-{1\over3}\frac{4N_{C}}{(4\pi)^{2}}a^{3}_{\mu}\partial^{2} a^{3\mu},\\
&&{\cal L}^{1+2+3+4}=\frac{4N_{C}}{(4\pi)^{2}}\sum^{\infty}_{k=1}
\frac{1}{(2k+1)(2k-1)}(\frac{m^{2}_{1}}{m^{2}_{2}})^{2k-2}
({g\over4})^{2}\sum^{2}_{i=1}A^{i}_{\mu}\partial^{2}A^{i\mu}+
\frac{4N_{C}}{3(4\pi)^{2}}a^{3}_{\mu}\partial^{2}a^{3\mu},\nonumber \\
&&{\cal L}^{5}=-{1\over2}\int\frac{d^{D}p}{(2\pi)^{D}}\sum^{\infty}
_{n=2}{1\over n}\frac{(-m^{2}_{2})^{n-2}}{(p^{2}+m^{2}_{1})^{n}}
\sum^{n-2}_{k=0}(n-1-k)Tr(\gamma\cdot v-\gamma\cdot a\gamma_{5})
\partial^{2}\tau^{k}_{3}(\gamma\cdot v+\gamma\cdot
a\gamma_{5})\tau^{n-2-k}	_{3}\nonumber \\
&&=-\frac{N_{C}}{(4\pi)^{2}}{D\over4}\Gamma(2-{D\over4})Tr(v_{\mu}\partial	^{2}v
^{\mu}+a_{\mu}\partial^{2}a^{\mu})\nonumber \\		
&&-\frac{4N_{C}}{(4\pi)^{2}}\sum^{\infty}_{k=2}\frac{1}{(2k-1)(2k-2)}
(\frac{m^{2}_{2}}{m^{2}_{1}})^{2k-2}({g\over4})^{2}A^{i}_{\mu}\partial^{2}
A^{i\mu}\nonumber \\
&&-\frac{2N_{C}}{(4\pi)^{2}}\sum^{\infty}_{k=2}\frac{1}{2k-2}(\frac{m^{2}_{2}}{
m^{2}_{1}})^{2k-2}\{v^{3}_{\mu}\partial^{2}v^{3\mu}+\omega_{\mu}
\partial^{2}\omega^{\mu}+a^{3}_{\mu}\partial^{2}a^{3\mu}\}\nonumber \\
&&+\frac{4N_{C}}{(4\pi)^{2}}\sum^{\infty}_{k=1}\frac{1}{2k-1}(\frac{m^{2}_{2}}{m
^{2}_{1}})^{2k-1}
\omega_{\mu}\partial^{2}v^{3\mu},\\
&&{\cal L}^{6}=2\int\frac{d^{D}p}{(2\pi)^{D}}\sum^{\infty}
_{n=3}{1\over n}\frac{(-m^{2}_{2})^{n-3}}{(p^{2}+m^{2}_{1})^{n}}\sum
^{n-3}_{k=0}\sum^{n-2-k}_{k_{1}=0}(n-2-k-k_{1})\nonumber \\
&&Trp\cdot(v+a\gamma_{5})\tau^{k
+k_{1}}_{3}\partial^{2}p\cdot(v+a\gamma_{5})\tau^{n-3-k-k_{1}}\nonumber \\
&&={2\over3}\frac{N_{C}}{(4\pi)^{2}}{D\over4}\Gamma(2-{D\over4})Tr(v_{\mu}
\partial^{2}v^{\mu}
+a_{\mu}\partial^{2}a^{\mu})\nonumber \\		
&&+\frac{8N_{C}}{(4\pi)^{2}}\sum^{\infty}_{k=2}\frac{1}{(2k+1)(2k-1)(2k-2)}
(\frac{m^{2}_{2}}{m^{2}_{1}})^{2k-2}({g\over4})^{2}A^{i}_{\mu}\partial^{2}
A^{i\mu}\nonumber \\
&&+\frac{4N_{C}}{3(4\pi)^{2}}\sum^{\infty}_{k=2}\frac{1}{2k-2}(\frac{m^{2}_{2}}{
m^{2}_{1}})^{2k-2}\{v^{3}_{\mu}\partial^{2}v^{3\mu}+\omega_{\mu}
\partial^{2}\omega^{\mu}+a^{3}_{\mu}\partial^{2}a^{3\mu}\}\nonumber \\
&&-\frac{8N_{C}}{3(4\pi)^{2}}\sum^{\infty}_{k=1}\frac{1}{2k-3}(\frac{m^{2}_{2}}
{m^{2}_{1}})^{2k-3}
\omega_{\mu}\partial^{2}v^{3\mu},\\
&&{\cal L}^{7}=-8\int\frac{d^{D}p}{(2\pi)^{D}}\sum^{\infty}
_{n=4}{1\over n}\frac{(-m^{2}_{2})^{n-4}}{(p^{2}+m^{2}_{1})^{n}}\sum
_{k,k_{1},k_{2}}(n-3-k-k_{1}-k_{2})\nonumber \\
&&Trp\cdot(v+a\gamma_{5})\tau^{k+k_{1}+k_{2}}
(p\cdot\partial)^{2}p\cdot(v+a\gamma_{5})\tau^{n-4-k-k_{1}-k_{2}}
\nonumber \\
&&=-{1\over3}\frac{N_{C}}{(4\pi)^{2}}{D\over4}\Gamma(2-{D\over4})Tr(v_{\mu}
\partial^{2}v^{\mu}-2\partial_{\mu}v^{\mu}\partial_{\nu}v^{\nu}
+a_{\mu}\partial^{2}a^{\mu}-2\partial_{\mu}a^{\mu}\partial_{\nu}a^{\nu})
\nonumber \\	
&&-\frac{4N_{C}}{(4\pi)^{2}}\sum^{\infty}_{k=3}\frac{1}{(2k-1)(2k-3)(2k-4)}
(\frac{m^{2}_{2}}{m^{2}_{1}})^{2k-2}({g\over4})^{2}(A^{i}_{\mu}\partial^{2}
A^{i\mu}-2\partial_{\mu}A^{i\mu}\partial_{\nu}A^{i\nu})\nonumber \\
&&-\frac{2N_{C}}{3(4\pi)^{2}}\sum^{\infty}_{k=3}\frac{1}{2k-4}(\frac{m^{2}_{2}}
{m^{2}_{1}})^{2k-4}\{v^{3}_{\mu}\partial^{2}v^{3\mu}-2\partial
_{\mu}v^{3\mu}\partial_{\nu}v^{3\nu}
+\omega_{\mu}\partial^{2}\omega^{\mu}\nonumber \\
&&-2\partial_{\mu}\omega^{\mu}\partial_{\nu}\omega^{\nu}
+a^{3}_{\mu}\partial^{2}a^{3\mu}-2\partial_{\mu}a^{3\mu}\partial_{\nu}
a^{3\nu}\}\nonumber \\
&&+\frac{4N_{C}}{3(4\pi)^{2}}\sum^{\infty}_{k=2}\frac{1}{2k-3}(\frac{m^{2}_{2}}
{m^{2}_{1}})^{2k-3}(
\omega_{\mu}\partial^{2}v^{3\mu}-2\partial_{\mu}\omega^{\mu}
\partial_{\nu}v^{3\nu}),\\
&&{\cal L}^{8}=2\int\frac{d^{D}p}{(2\pi)^{D}}\sum^{\infty}
_{n=3}{1\over n}\frac{(-m^{2}_{2})^{n-3}}{(p^{2}+m^{2}_{1})^{n}}\sum
^{n-3}_{k=0}\sum^{n-3-k}_{k_{1}=0}(n-2-k-k_{1})\nonumber \\
&&Trp\cdot(v+a\gamma_{5})\tau
^{k+k_{1}}_{3}p\cdot\partial\gamma\cdot\partial(\gamma\cdot v+\gamma\cdot a
\gamma_{5})\tau^{n-3-k-k_{1}},\nonumber  \\	
&&={2\over3}\frac{N_{C}}{(4\pi)^{2}}{D\over4}\Gamma(2-{D\over4})Tr(v^{\mu}
\partial_{\mu\nu}v^{\nu}+a^{\mu}\partial_{\mu\nu}a^{\nu})\nonumber \\
&&+\frac{8N_{C}}{(4\pi)^{2}}\sum^{\infty}_{k=2}\frac{1}{(2k+1)(2k-1)(2k-2)}
(\frac{m^{2}_{2}}{m^{2}_{1}})^{2k-2}({g\over4})^{2}(A^{i\mu}\partial
_{\mu\nu}A^{i\nu})\nonumber \\
&&+\frac{4N_{C}}{3(4\pi)^{2}}\sum^{\infty}_{k=2}\frac{1}{2k-2}(\frac{m^{2}_{2}}
{m^{2}_{1}})^{2k-2}\{v^{3\mu}\partial_{\mu\nu}
v^{3\nu}
+\omega^{\mu}\partial_{\mu\nu}\omega^{\nu}+a^{3\mu}\partial_{\mu\nu}
a^{3\nu})\nonumber \\
&&-\frac{8N_{C}}{3(4\pi)^{2}}\sum^{\infty}_{k=2}\frac{1}{2k-3}(\frac{m^{2}_{2}}
{m^{2}_{1}})^{2k-3}
\omega^{\mu}\partial_{\mu\nu}v^{3\nu},\\
&&{\cal L}^{9}=2\int\frac{d^{D}p}{(2\pi)^{D}}\sum^{\infty}
_{n=3}{1\over n}\frac{(-m^{2}_{2})^{n-3}}{(p^{2}+m^{2}_{1})^{n}}\sum
^{n-3}_{k=0}\sum^{n-3-k}_{k_{1}=0}(n-2-k-k_{1})\nonumber \\
&&Tr
(\gamma\cdot v-\gamma\cdot a\gamma_{5})\gamma\cdot\partial p\cdot\partial
\tau
^{k+k_{1}}_{3}p\cdot(\gamma\cdot v+\gamma\cdot a
\gamma_{5})\tau^{n-3-k-k_{1}},\nonumber  \\	
&&={\cal L}^{8}.
\end{eqnarray}

Taking other two generations of quarks and three generations of
leptons into account in Eqs.(24-32) and adding them  together,
in Minkowski space the kinetic terms are obtained
\begin{eqnarray}
\lefteqn{{\cal L}_{K}=-{1\over4}\sum_{i=1,2}
(\partial_{\mu}A^{i}_{\nu}-
\partial_{\nu}A^{i}_{\mu})^{2}\{1+{1\over(4\pi)^{2}}g^{2}
\sum_{q,l}N[{D\over12}
\Gamma(2-{D\over2})(4\pi)^{{\epsilon\over2}}({\mu^{2}\over m^{2}_{1}})
^{{\epsilon\over2}}-{1\over6}+f_{1}]\}}\nonumber \\
&&-{1\over4}(\partial_{\mu}A^{3}_{\nu}-\partial_{\nu}A^{3}_{\mu})^{2}
\{1+\frac{1}{(4\pi)^{2}}g^{2}\sum_{q,l}N
[{D\over12}\Gamma(2-{D\over2})
(4\pi)^{{\epsilon\over2}}({\mu^{2}\over m^{2}_{1}})
^{{\epsilon\over2}}-{1\over6}+{1\over6}f_{2}]\}\nonumber \\
&&-{1\over4}(\partial_{\mu}B_{\nu}-\partial_{\nu}B_{\mu})^{2}
\{1+\frac{N_{C}}{(4\pi)^{2}}g'^{2}\sum_{q}[{11\over9}
{D\over12}\Gamma(2-{D\over2})
(4\pi)^{{\epsilon\over2}}({\mu^{2}\over m^{2}_{1}})
^{{\epsilon\over2}}-{1\over6}+{11\over54}f_{2}-{1\over18}f_{3}
]\}\nonumber \\
&&+\frac{1}{(4\pi)^{2}}g'^{2}\sum_{l}[
{D\over4}\Gamma(2-{D\over2})
(4\pi)^{{\epsilon\over2}}({\mu^{2}\over m^{2}_{1}})
^{{\epsilon\over2}}-{1\over6}+{1\over2}f_{2}+{1\over6}f_{3}
]\}\nonumber \\
&&-\frac{1}{(4\pi)^{2}}{gg'\over12}(\partial_{\mu}A^{3}_{\nu}
-\partial_{\nu}A^{3}_{\mu})(\partial^{\mu}B^{\nu}-\partial^{\nu}B^{\mu})
\{N_{G}-{2\over3}N_{C}\sum_{q}f_{3}+2\sum_{l}f_{3}\}\nonumber \\
&&+\frac{1}{(4\pi)^{2}}N_{G}
{g^{2}\over12}\sum_{i=1,2}(\partial^{\mu}A^{i}_{\mu})
^{2}+\frac{1}{(4\pi)^{2}}N_{G}
{1\over12}(g^{2}+g'^{2})(\partial_{\mu}
Z^{\mu})^{2},
\end{eqnarray}
where $\sum_{q}$ and $\sum_{l}$ stand for summations of
generations of quarks and leptons respectively,
\(N=N_{C}\) for q and \(N=1\) for l, \(N_{G}=3N_{C}+3\),
x depends on fermion generation and is defined in Eq.(22),
\begin{eqnarray}
\lefteqn{f_{1}={4\over9}-{1\over6x}
-{1\over6}ln(1-x)
+{1\over4\sqrt{x}}({1\over3x}-1)
ln\frac{1+\sqrt{x}}{1-\sqrt{x}},}\nonumber \\
&&f_{2}=-ln(1-x),\;\;\;
f_{3}=
{1\over2}{1\over\sqrt{x}}ln\frac{1+\sqrt{x}}
{1-\sqrt{x}}.
\end{eqnarray}
Following results are obtained from Eq.(33)
\begin{enumerate}
\item The boson fields and the coupling constants g and g'
have to be redefined by multiplicative renormalization
\begin{eqnarray}
A^{1,2}_{\mu}\rightarrow Z^{{1\over2}}_{1}A^{1,2}_{\mu},
A^{3}_{\mu}\rightarrow Z^{{1\over2}}_{3}A^{3}_{\mu},
B_{\mu}\rightarrow Z^{{1\over2}}_{B}B{\mu},\nonumber \\
g_{1}=Z^{-{1\over2}}_{1}g,g_{3}=Z^{-{1\over2}}_{3}g,
g_{B}=Z^{-{1\over2}}_{B}g',
\end{eqnarray}
where
\begin{eqnarray}
\lefteqn{Z_{1}=
1+{1\over(4\pi)^{2}}g^{2}\sum_{q,l}N
[{D\over12}
\Gamma(2-{D\over2})(4\pi)^{{\epsilon\over2}}
({\mu^{2}\over m^{2}_{1}})
^{{\epsilon\over2}}-{1\over6}+f_{1}]}\nonumber \\
&&Z_{3}=1+\frac{1}{(4\pi)^{2}}g^{2}\sum_{q,l}N[
{D\over12}\Gamma(2-{D\over2})
(4\pi)^{{\epsilon\over2}}({\mu^{2}\over m^{2}_{1}})
^{{\epsilon\over2}}-{1\over6}+{1\over6}f_{2}]\nonumber \\
&&Z_{B}=1+\frac{N_{C}}{(4\pi)^{2}}g'^{2}\sum_{q}[{11\over9}
{D\over12}\Gamma(2-{D\over2})
(4\pi)^{{\epsilon\over2}}({\mu^{2}\over m^{2}_{1}})
^{{\epsilon\over2}}-{1\over6}+{11\over54}f_{2}-{1\over18}f_{3}
]\nonumber \\
&&+\frac{1}{(4\pi)^{2}}g'^{2}\sum_{l}[
{D\over4}\Gamma(2-{D\over2})
(4\pi)^{{\epsilon\over2}}({\mu^{2}\over m^{2}_{1}})
^{{\epsilon\over2}}-{1\over6}+{1\over2}f_{2}+{1\over6}f_{3}
].
\end{eqnarray}
The divergent terms in $Z_{1}$ and $Z_{3}$ are the same.
\item There is a crossing term between
$A^{3}_{\mu}$ and $B_{\mu}$, which is written as
\begin{eqnarray}
\lefteqn{g_{3}g_{B}(\partial_{\mu}A^{3}_{\nu}-\partial_{\nu}A^{3}_{\mu})
(\partial_{\mu}B_{\nu}-\partial_{\nu}B_{\mu})
=e^{2}(\partial_{\mu}A_{\nu}-\partial_{\nu}A_{\mu})^{2}
-e^{2}(\partial_{\mu}Z_{\nu}-\partial_{\nu}Z_{\mu})^{2}}
\nonumber \\
&&+e(g_{3}cos\theta_{W}-g_{B}sin\theta_{W})
(\partial_{\mu}A_{\nu}-\partial_{\nu}A_{\mu})
(\partial_{\mu}Z_{\nu}-\partial_{\nu}Z_{\mu}),
\end{eqnarray}
where \(sin\theta_{W}=\frac{g_{B}}{\sqrt{g^{2}_{B}+g^{2}_{3}}}\),
\(cos\theta_{W}=\frac{g_{3}}{\sqrt{g^{2}_{B}+g^{2}_{3}}}\),
and \(e=\frac{g_{3}g_{B}}{\sqrt{g^{2}_{B}+g^{2}_{3}}}\).
Therefore, the photon and the Z fields are needed to be renormalized
again
\begin{equation}
(1+{\alpha\over4\pi}f_{4})^{{1\over2}}A_{\mu}\rightarrow A_{\mu},\;\;\;
(1-{\alpha\over4\pi}f_{4})^{{1\over2}}Z_{\mu}\rightarrow Z_{\mu},
\end{equation}
where
\[f_{4}={1\over3}N_{G}-{2\over3}\sum_{q}f_{3}+{2\over3}\sum_{l}f_{3}.\]
After these renormalizations(35,39),
${\cal L}_{K}$(33) is rewritten as
\begin{eqnarray}
\lefteqn{{\cal L}_{K}=
-{1\over 4}(\partial_{\mu}A_{\nu}-\partial_{\nu}
A_{\mu})^{2}
-{1\over 4}\sum_{i=1,2}(\partial_{\mu}A^{i}_{\nu}-\partial_{\nu}
A^{i}_{\mu})^{2}
-{1\over 4}(\partial_{\mu}Z_{\nu}-\partial_{\nu}
Z_{\mu})^{2}}\nonumber \\
&&-{1\over4}\frac{\alpha}{4\pi}
({g_{3}\over g_{B}}-{g_{B}\over g_{3}})
(1-\frac{\alpha^{2}}{(4\pi)
^{2}}f^{2}_{4})^{-{1\over2}}f_{4}
(\partial_{\mu}A_{\nu}-
\partial_{\nu}A_{\mu})(\partial^{\mu}Z^{\nu}-\partial^{\nu}Z^{\mu})
\nonumber \\
&&-\frac{N_{G}}{(4\pi)^{2}}
{g^{2}\over12}(\partial^{\mu}A^{i}_{\mu})
^{2}-\frac{N_{G}}{(4\pi)^{2}}(1-{\alpha\over4\pi}f_{4})^{-1}
{1\over12}(g^{2}_{3}+g^{2}_{B})(\partial_{\mu}
Z^{\mu})^{2}.
\end{eqnarray}
\item
The interaction between photon and Z boson is predicted in Eq.(39).
\item For very small neutrino masses it is derived from Eq.(34)
\begin{equation}
f_{3}=-ln({m_{\nu}\over m_{l}}).
\end{equation}
Therefore, if neutrino is massless
$f_{4}$ is logarithmic divergent. This divergence is in contradiction
with that the physical coupling between photon and z boson must be
finite.
Therefore, this theory requires massive neutrinos.
\end{enumerate}
\section{Gauge fixing term}
Before the spontaneous symmetry breaking
the standard model is gauge invariant.
Gauge fixing terms can be chosen artificially.
In the Lagrangian(2) the strengths
of the boson fields have the structure of the Yang-Mills fields. However, W and
Z are massive . Eq.(39) shows that
fixed gauge fixing terms of W- and Z- fields are dynamically generated
\begin{eqnarray}
-{1\over4}\xi_{W}(\partial^{\mu}A^{i}_{\mu})^{2},\;\;\;
-{1\over4}\xi_{Z}(\partial^{\mu}Z_{\mu})^{2},\nonumber \\
\xi_{W}=\frac{N_{G}}{(4\pi)^{2}}{g^{2}_{1}\over3},\;\;\;
\xi_{Z}=\frac{N_{G}}{(4\pi)^{2}}(1-{\alpha\over4\pi}f_{4})^{-1}
{1\over3}(g^{2}_{3}+g^{2}_{B}).
\end{eqnarray}
The propagator of W field is derived from Eq.(39)
\begin{equation}
\frac{i}{q^{2}-m^{2}_{W}}\{-g_{\mu\nu}+\frac{q_{\mu}q_{\nu}}{q^{2}}
\}-\frac{i}{\xi_{W}q^{2}-m^{2}_{W}}\frac{q_{\mu}q_{\nu}}{q^{2}}.
\end{equation}
Changing the index W to Z in Eq.(42),
the propagator of Z boson field
is obtained. Obviously, due to the gauge fixing terms the propagators
of W an Z bosons do not affect the renormalizability of the theory(2).
It is necessary to point out that the W and the Z fields are massive
and no longer gauge fields.
The "gauge fixing"
terms of W and Z bosons(41) are derived from this theory and they
are not obtained by choosing gauge.
\section{Theoretical values of $m_{W}$ and $m_{Z}$}
Now we can study the values of the masses of W and Z bosons.
After the renormalizations(35,38) there are still divergences in the
mass formulas of $m_{W}$ and $m_{Z}$(22). The boson fields are
already renormalized and the kinetic terms of the boson fields
are already in the standard form. Therefore, the divergences
in the formulas of $m_{W}$ and $m_{Z}$
cannot be absorbed by the boson fields. On the other hand,
the divergences in Eq.(22) are fermion mass dependent, while
the coupling constants should be the same for all fermion
generations. It is difficult that these divergences are absorbed
by the coupling constants. In Eq.(22) the fermion masses are bare
physical quantities. It is reasonable to redefine the fermion
masses by multiplicative renormalization
\begin{eqnarray}
Z_{m}m^{2}_{1}=m^{2}_{1,P},\nonumber \\
Z_{m}=\frac{N}{(4\pi)^{2}}\{N_{G}{D\over4}\Gamma(2-{D\over2})
(4\pi)^{{\epsilon\over2}}({\mu^{2}\over m^{2}_{1}})
^{{\epsilon\over2}}+{1\over2}[1-ln(1-x)-
(1+{1\over x}){\sqrt{x}\over2}
ln\frac{1+\sqrt{x}}{1-\sqrt{x}}]\},
\end{eqnarray}
for each generation of fermions. The index "P" is omitted in the rest
of the paper.
Now the mass of W boson is obtained from Eq.(22)
\begin{equation}
m^{2}_{W}={1\over2}g^{2}\{m^{2}_{t}+m^{2}_{b}+m^{2}_{c}+m^{2}_{s}
+m^{2}_{u}+m^{2}_{d}+m^{2}_{\nu_{e}}+m^{2}_{e}
+m^{2}_{\nu_{\mu}}+m^{2}_{\mu}+m^{2}_{\nu_{\tau}}+m^{2}_{\tau}\}
\end{equation}
Obviously, the top quark mass dominates the $m_{W}$
\begin{equation}
m_{W}={g\over\sqrt{2}}m_{t}.
\end{equation}
Using the values \(g=0.642\) and \(m_{t}=180\pm12 GeV\)[6], it is found
\begin{equation}
m_{W}=81.71(1\pm0.067) GeV,
\end{equation}
which is in excellent agreement with data $80.33\pm0.15$GeV[6].
The Fermi coupling constant is derived from eq.(45)
\[G_{F}=\frac{1}{2\sqrt{2}m^{2}_{t}}=0.96\times10^{-5}m^{-2}_{N},\]
where \(m_{t}=180GeV\) is taken.

Using Eqs.(22,35,38),
the mass formula of the Z boson
is written as
\begin{equation}
m^{2}_{Z}=\rho m^{2}_{W}(1+{g^{2}_{B}\over g^{2}_{1}}),
\end{equation}
where
\begin{eqnarray}
\lefteqn{\rho=(1-{\alpha\over4\pi}f_{4})^{-1}\sum_{q,l}N
\{{D\over4}
\Gamma(2-{D\over2})(4\pi)
^{{\epsilon\over2}}({\mu^{2}\over m^{2}_{1}})
^{{\epsilon\over2}}-{1\over2}[ln(1-x)+\sqrt{x}
ln\frac{1+\sqrt{x}}{1-\sqrt{x}}]\}}
\nonumber \\
&&/\sum_{q,l}N\{{D\over4}
\Gamma(2-{D\over2})(4\pi{\mu^{2}\over m^{2}_{1}})
^{{\epsilon\over2}}+{1\over2}[1-ln(1-x)-
(1+{1\over x}){\sqrt{x}\over2}
ln\frac{1+\sqrt{x}}{1-\sqrt{x}}]\}
\frac{g^{2}_{3}+g^{2}_{B}}
{g^{2}_{1}+g^{2}_{B}}.
\end{eqnarray}
Comparing
with the infinites in Eqs.(35,36,48), the finite terms can be
ignored in Eqs.(35,36,48). We have
\begin{equation}
g_{1}=g_{3}\equiv g_{A},\;
m^{2}_{Z}=\rho m^{2}_{W}/cos^{2}\theta_{W},\;
cos\theta_{W}=g_{A}/\sqrt{g^{2}_{A}+g^{2}_{B}},\;
\rho=(1-{\alpha\over4\pi}f_{4})^{-1},
\end{equation}
where $g_{A}$ and $g_{B}$ are g and g' of the GWS model respectively.
The finiteness of the $\rho$ factor requires a finite $f_{4}$.
Once again massive neutrinos are required.
Due to the smallness of the factor ${\alpha\over4\pi}$
in the reasonable ranges of the quark masses
and the upper
limits of neutrino masses we expect
\begin{equation}
\rho\simeq 1.
\end{equation}
Therefore,
\begin{equation}
m_{Z}=m_{W}/cos\theta_{W}
\end{equation}
is a good approximation. Eq.(51) is the prediction of the standard model.

Introduction of a cut-off leads to
\begin{equation}
{D\over 4}
\Gamma(2-{D\over2})(4\pi)^{{\epsilon\over2}}({\mu^{2}\over m^{2}_{1}})
^{{\epsilon\over2}}\rightarrow
ln(1+{\Lambda^{2}\over
m^{2}_{1}})-1+\frac{1}{1+{\Lambda^{2}\over m^{2}_{1}}}.
\end{equation}
Taking $\Lambda\rightarrow\infty$,
Eq.(49) is obtained.
On the other hand,
the cut-off might be estimated by the value of the $\rho$
factor. The cut-off can be
considered as
the energy scale of unified electroweak theory.
\section{Summary}
In the Lagrangian(2) the boson fields are elementary fields. The Lagrangian
of boson fields, the couplings between the fermions and the bosons, and
the fermion mass terms are the same as in the standard model. The boson
fields are different from photon and gluons, they have both vector and
axial-vector components. It shows in this paper that
axial-vector field is very different from vector field. A new dynamical
symmetry breaking caused by the combination of the axial-vector component
and fermion mass is found. In this theory there are explicit symmetry
breaking by fermion masses and dynamical symmetry breaking. The charged
boson, W, gains mass from both the symmetry breaking, while  the mass of
the neutral boson, Z, originates in the dynamical symmetry breaking only.
Due to the U(1) symmetry and the vector nature of photon the photon remains
massless. Upon the scheme of multiplicative renormalization the values of
$m_{W}$ and $m_{Z}$ are determined. They are in excellent agreement with
data. After W and Z gain masses the theory is no longer gauge invariant.
Two gauge
fixing terms for W and Z are dynamically generated respectively.
The propagators of W- and Z-fields have no problem for renormalization.
On the other hand,
The finiteness
of $m_{Z}$, the $\gamma-Z$ coupling, and the fixed gauging fixing
terms require massive neutrinos.

In this paper[8] the path integral is used to derive the Lagrangian of boson
fields without any additional assumption. The results are rigorous.
The Lagrangian(2) is the one of the standard model without Higgs.
Therefore, the successes of the
standard model are kept.
$m_{W}$, $m_{Z}$, and gauge fixing terms are dynamically generated.
This theory is equivalent to the standard model without Higgs.

This research was partially
supported by DOE Grant No. DE-91ER75661.

\end{document}